\documentclass[prc,showpacs,preprintnumbers,amsmath,amssymb,superscriptaddress,floatfix,nofootinbib]{revtex4}
\usepackage{amsmath,amssymb}
\usepackage{epsfig}
\usepackage{color}
\usepackage{bm}
\usepackage{epstopdf}

\newcommand{\X}{X(3872)}
\newcommand{\DD}{D^0\bar{D}^0}

\makeatletter
\def\slashchar#1{{\mathpalette\c@ncel{#1}}} 
\makeatother

\begin{document}

\title{Hidden charm and bottom molecular states}
 
\author{F.-K. Guo} 
\affiliation{Helmholtz-Institut f\"ur Strahlen- und
             Kernphysik and Bethe Center for Theoretical Physics, 
             Universit\"at Bonn,  D-53115 Bonn, Germany}
\author{C. Hidalgo-Duque}
\affiliation{  Instituto de F\'isica Corpuscular (IFIC),
             Centro Mixto CSIC-Universidad de Valencia,
             Institutos de Investigaci\'on de Paterna,
             Aptd. 22085, E-46071 Valencia, Spain}
\author{J. Nieves} 
\affiliation{  Instituto de F\'isica Corpuscular (IFIC),
             Centro Mixto CSIC-Universidad de Valencia,
             Institutos de Investigaci\'on de Paterna,
             Aptd. 22085, E-46071 Valencia, Spain}
\author{A. Ozpineci}
\affiliation{Middle East Technical University - Department of Physics
TR-06531 Ankara, Turkey  }

\author{M. Pav\'on Valderrama}
\affiliation{Institut de Physique Nucl\'eaire,
             Universit\'e Paris-Sud, IN2P3/CNRS,
             F-91406 Orsay Cedex, France}

\begin{abstract}
We investigate heavy quark  symmetries for heavy light meson-antimeson
systems in a contact-range 
effective field theory. In the SU(3) light flavor limit, the leading
order Lagrangian respecting heavy quark spin symmetry contains four
independent counter-terms. Neglecting $1/m_Q$ corrections, three of
these low energy constants can be determined 
by theorizing a molecular description of the $X(3872)$ and $Z_b(10610)$
states. Thus, we can predict new hadronic molecules, in particular the isovector
charmonium partners of the $Z_b(10610)$ and the $Z_b(10650)$
states. We also discuss hadron molecules
composed of a heavy meson and a doubly-heavy baryon, which would be
related to the  heavy meson-antimeson molecules thanks to the
  heavy antiquark-diquark symmetry. Finally, we also study the $\X \to
  \DD\pi^0$ decay, which is not only sensitive to the short distance part of the $\X$
molecular wave function, as the $J/\psi\pi\pi$ and $J/\psi3\pi$ $\X$
decay modes are, but it is also affected by the long-distance
structure of the resonance. Furthermore, this decay
might provide some information on the interaction between
the $D\bar D$ charm mesons.

\end{abstract}
\maketitle

\section{Heavy meson molecules}

The recent  discoveries of exotic  heavy quarkonium  states revived
old expectations on the possible existence of molecular (loosely bound
mesonic colour singlets) states~\cite{Voloshin:1976ap,DeRujula:1976qd}. 
The most likely candidates are the $X(3872)$ resonance~\cite{Choi:2003ue} and the isovector $Z_b(10610)$ and
$Z_b(10650)$ states~\cite{Belle:2011aa,Adachi:2012im}. The proximity
of the $\X$ to the $D \bar D^{*0}$ threshold and its decay properties have
led to the general acceptance that it is a weakly bound state, with quantum numbers $J^{PC} =
1^{++}$~\cite{Aaij:2013zoa}, generated from   $D^0\bar
D^{*0}$, $D^+\bar D^{*-}$ coupled channel
interactions~\cite{Gamermann:2009uq}\footnote{When we refer to $D^0
  \bar D^{*0} , D^+ D^{*-}$, or in general $D \bar D^*$, we are actually referring to the combination of
  these states with their charge conjugate ones in order to form a
  state with well-defined C-parity.}. 
Heavy quark
symmetries deduced from QCD provide an adequate framework to study
these systems. Thus, heavy quark spin symmetry (HQSS) implies that
molecular states should appear in HQSS multiplets, while from heavy flavor
symmetry (HFS), similarities in the bottom and charm spectra might be
expected. Indeed, combining both HQSS and HFS, various partners of the $\X$ and
the isovector $Z_b'^s$ states can be
predicted~\cite{Guo:2009id,Bondar:2011ev,Voloshin:2011qa,Mehen:2011yh,Nieves:2011zz,Nieves:2012tt,HidalgoDuque:2012pq,Guo:2013sya}.

\begin{table}[h]
\caption{Heavy meson--heavy meson combinations
  having the same contact term as the $X(3872)$ and $Z_b(10610)$, and
  the predictions of the masses, which are understood to
  correspond to bound states except if we write ``V'' in parenthesis
  for denoting a virtual state. $\dagger$: increasing the strength of the
  potential to account for the various uncertainties, the virtual pole
  evolves into a bound state. Masses are given  (MeV units) for two
  UV regulators. For further details see \cite{Guo:2013sya}.}
\label{tab:1}       
\begin{tabular}{ c c c  c c}
       $I(J^{PC})$ & States & $M$ ($\Lambda=0.5$ GeV) &
       $M$ ($\Lambda=1$ GeV) & Measurements
       \\\hline\\
        $0(1^{++})$ & $D\bar D^*$ & 3871.68 (input) &  3871.68 (input) &
       $3871.68\pm0.17$~\cite{PDG}  \\
                        $0(2^{++})$ & $D^*\bar D^*$ & $4012^{+4}_{-5}$  &  $4012^{+5}_{-12}$ & ?\\
        $0(1^{++})$ & $B\bar B^*$ & $10580^{+9}_{-8}$ &  $10539^{+25}_{-27}$ & ?\\
                        $0(2^{++})$ & $B^*\bar B^*$ & $10626^{+8}_{-9}$ & $10584^{+25}_{-27}$ & ?\\
                        $0(2^{+})$ & $D^*B^*$ & $7322^{+6}_{-7}$ &
       $7308^{+16}_{-20}$ & ?\\ \hline \\
        $1(1^{+-})$ & $B\bar B^*$ & $10602.4 \pm 2.0$ (input) & $10602.4 \pm 2.0$ (input) &
       $10607.2\pm2.0$~\cite{Belle:2011aa} \\
        & & &  & $10597\pm9$~\cite{Adachi:2012cx}\\
                        $1(1^{+-})$ & $B^*\bar B^*$ & $10648.1 \pm 2.1
       $ &  $10648.1 ^{+2.1}_{-2.5}$ & $10652.2\pm1.5$~\cite{Belle:2011aa} \\
        & & &  & $10649\pm12$~\cite{Adachi:2012cx}\\
                        $1(1^{+-})$ & $D\bar D^*$ & $3871^{+4}_{-12}$ (V) &  $3837_{-35}^{+17}$ (V) & $3899.0 \pm 3.6 \pm 4.9$~\cite{BESIII:2013} \\
        & & & & $3894.5 \pm 6.6 \pm 4.5$~\cite{Liu:2013xoa}\\
                        $1(1^{+-})$ & $D^*\bar D^*$ & $4013^{+4}_{-11}$ (V) &  $3983_{-32}^{+17}$ (V) & ? \\
                        $1(1^{+})$ & $D^*B^*$ & $7333.6^{\dagger}_{-4.2}$ (V) & $7328^{+5}_{-14}$ (V) & ?\\
   \end{tabular}
\end{table}

Actually, HQSS heavily constrains the low-energy interactions among
heavy hadrons \cite{Bondar:2011ev,Mehen:2011yh,Nieves:2012tt,HidalgoDuque:2012pq,AlFiky:2005jd}. As long as the hadrons are not
too tightly bound, they will not probe the specific details of the
interaction binding them at short distances. Moreover, each of the
constituent heavy hadrons will be unable to see the internal structure
of the other heavy hadron. This separation of scales can be used to
formulate an effective field theory (EFT) description of hadronic
molecules~\cite{Mehen:2011yh,Nieves:2012tt} compatible with the
approximate nature of HQSS.  At very low  energies, the leading order (LO)
interaction between pseudoscalar 
and vector charmed  ($D^0, D^+, D^{*0}, D^{*+}$) and anti-charmed
($\bar D^0, D^-, \bar D^{*0}, D^{*-}$) mesons\footnote{The discussion
  runs in parallel for the bottom sector.} can be
described just in terms of a contact-range potential, which is constrained by HQSS~\cite{Nieves:2012tt,HidalgoDuque:2012pq,Guo:2013sya}. Pion exchange and
particle coupled-channel\footnote{We do not refer to charge channels, 
but rather to the mixing among the $D\bar D$, $D\bar D^*$, $D^*\bar
D^*$ pairs in a given $IJC$ (isospin, spin and charge conjugation)
sector.} effects turn out to be sub-leading~\cite{Nieves:2012tt,Valderrama:2012jv}.

The LO Lagrangian contains four independent
terms in the SU(3) flavor limit~\cite{HidalgoDuque:2012pq}, which
strength is set 
by two isoscalar $C_{0A}$ and $C_{0B}$ and two isovector $C_{1A}$ and
$C_{1B}$ low energy constants (LEC's). The (contact) interaction
potential is used as kernel of a two body elastic Lippmann-Swinger
equation (LSE). The LSE shows
an ill-defined ultraviolet (UV) behaviour, and it requires a
regularization and renormalization procedure (see
Refs.~\cite{HidalgoDuque:2012pq,Guo:2013sya} for details). 
The LSE non-perturbative re-summation 
restores elastic unitarity and provides  a non-analytical structure of
the scattering amplitudes. Bound states ($D^{(*)}\bar D^{(*)}$ or $B^{(*)}\bar
B^{(*)}$ molecules) correspond to poles of the $T$-matrix
below threshold on the real axis in the first Riemann sheet of the
complex energy, while  virtual states, that if located near threshold
might be relevant, show up in the second Riemann sheet. 

Two combinations of the LEC's can be obtained from the properties of
the $\X$ resonance\footnote{Mass and the isospin violating
ratio of the decay amplitudes for the $X(3872) \to J/\psi\pi\pi$ and $X(3872) \to 
J/\psi\pi\pi\pi$ transitions, $R_{X(3872)}=0.26\pm 0.07$ deduced in
~\cite{Hanhart:2011tn} using the experimental  ratio $\mathcal{B}_{X} = \Gamma
\left[X \to J / \Psi \,\rho\, \right]/\Gamma \left[X \to J / \Psi
  \,\omega\,\right] = 1.3 \pm 0.5$ \cite{Choi:2011fc}.}, assuming that it is  a $\left( D\bar D^*-D^* \bar
D\right)/\sqrt{2}$ bound state. The isospin properties of the
$\X$ molecule are mainly determined by its mass, which is only
few tens of keV below the $D^0\bar  D^{*0}$ threshold,  making 
relevant the around 8 MeV difference between the threshold of the neutral and 
of the charged ($D^+ D^{*-}$) channels~\cite{Gamermann:2009uq,HidalgoDuque:2012pq}.

Assuming HFS, a third independent combination of LEC's is fixed from the isovector
$Z_b(10610)$ resonance, described  as $\left(B\bar B^*+B^* \bar
B\right)/\sqrt{2}$ molecular state. Note, HQSS predicts the
interaction of the $B^*\bar B^*$ system with  $I=1, J^{PC}=1^{+-}$
quantum numbers to be identical to that of the $B\bar B^*$ pair in the
$Z_b(10610)$ sector. Thus, HQSS naturally explains the approximate
degeneracy of the $Z_b(10610)$ and $Z_b(10650)$ resonances~\cite{Belle:2011aa,Adachi:2012im}. 

There are various $D^{(*)}\bar D^{(*)}$, $B^{(*)}\bar
B^{(*)}$ and $D^{(*)}\bar B^{(*)}$  sectors where the interaction is
completely fixed by the the three linear combinations of LEC's
obtained from the $\X$ and the $Z_b(10610)$ resonances, which allows
us to make predictions on the existence of additional molecular states, by solving the LSE as previously
commented. Some of these approximate predictions from \cite{Guo:2013sya} are collected in
Table~\ref{tab:1}.

\section{Triply heavy pentaquarks}

\begin{table}
\caption{Doubly-heavy baryon--heavy meson
molecules masses.
The error in
the masses of the isoscalar states is a consequence of the approximate nature of HADS.
For the isovector states, different error sources have been taken into account: the
uncertainty in the $Z_b$ binding, in the isospin breaking decays of the $X$
and in the HADS breaking. For simplicity, we only show an unique error obtained
by adding in quadratures all the previous ones.
$M_{\rm th}$ represents the threshold, and all masses are given in units
of MeV.  When we decrease the
strength of the potential to account for the various uncertainties, in some 
cases (marked with $\dagger$ in the table)  the bound state pole reaches the 
threshold and the state becomes virtual. The cases with a virtual state pole 
at the central value are marked by [V], for which $\dagger\dagger$ 
means that the pole 
evolves into a bound state one and N/A means that the pole is far from the 
threshold with a momentum larger than 1~GeV so that it is both undetectable and 
beyond the EFT range. For further details see \cite{Guo:2013xga}.}
\label{tab:2}       
\begin{tabular}{l c c c c c}
       State & $I(J^{P})$ & $V^{\rm LO}$ & Thresholds &
       M ($\Lambda=0.5$ GeV) & M($\Lambda=1$ GeV) \\ \hline
       $\Xi_{cc}^* D^*$ & $0({\frac{5}{2}}^-)$ & $C_{0a}+C_{0b}$ & $5715$
       & $\left( M_{\rm th} - 10\right)^{+10}_{-15}$ & $\left(M_{\rm th} - 
19\right)^{\dagger}_{-44}$ \\
       $\Xi_{cc}^* \bar{B}^*$ & $0({\frac{5}{2}}^-)$ & $C_{0a}+C_{0b}$ &
       $9031$ & $\left(M_{\rm th} - 21\right)^{+16}_{-19}$ & $\left(M_{\rm th} - 
53\right)^{+45}_{-59}$ \\
       $\Xi_{bb}^* D^*$ & $0({\frac{5}{2}}^-)$ & $C_{0a}+C_{0b}$ & $12160$
       & $\left(M_{\rm th} - 15\right)^{+9}_{-11}$ & $\left(M_{\rm th} - 
35\right)^{+25}_{-31}$ \\
       $\Xi_{bb}^* \bar{B}^*$ & $0({\frac{5}{2}}^-)$
       & $C_{0a}+C_{0b}$ & $15476$
       & $\left(M_{\rm th} - 29\right)^{+12}_{-13}$ & $\left(M_{\rm th} - 
83\right)^{+38}_{-40}$ \\
       $\Xi_{bc}' D^*$ & $0({\frac{3}{2}}^-)$ & $C_{0a}+C_{0b}$ & $8967$
       & $\left(M_{\rm th} - 14\right)^{+11}_{-13}$ & $\left(M_{\rm th} - 
30\right)^{+27}_{-40}$ \\
       $\Xi_{bc}' \bar{B}^*$ & $0({\frac{3}{2}}^-)$
       & $C_{0a}+C_{0b}$ & $12283$
       & $\left(M_{\rm th} - 27\right)^{+15}_{-16}$ & $\left(M_{\rm th} - 
74\right)^{+45}_{-51}$ \\
$\Xi_{bc}^* D^*$ & $0({\frac{5}{2}}^-)$ & $C_{0a}+C_{0b}$ & $9005$
& $\left(M_{\rm th} - 14\right)^{+11}_{-13}$ & $\left(M_{\rm th} -
30\right)^{+27}_{-40}$ \\
$\Xi_{bc}^* \bar{B}^*$ & $0({\frac{5}{2}}^-)$ & $C_{0a}+C_{0b}$ & $12321$
& $\left(M_{\rm th} -27\right)^{+15}_{-16}$ & $\left(M_{\rm th} 
-74\right)^{+46}_{-51}$ \\\hline
       $\Xi_{bb}\bar{B}$ & $1({\frac{1}{2}}^-)$ & $C_{1a}$ & $15406$
       & $\left(M_{\rm th} - 0.3\right)_{-2.5}^{\dagger}$ &
       $\left(M_{\rm th} -  12\right)^{+11}_{-15}$ \\
       $\Xi_{bb}\bar{B}^*$ & $1({\frac{1}{2}}^-)$ &
       $C_{1a}+\frac{2}{3}\,C_{1b}$ & $15452$
       & ($M_{\rm th}-0.9$)[V]$_{\dagger\dagger}^{\rm N/A}$
       & $\left(M_{\rm th} - 16\right)^{+14}_{-17}$ \\
       $\Xi_{bb}\bar{B}^*$ & $1({\frac{3}{2}}^-)$
       & $C_{1a}-\frac{1}{3}\,C_{1b}$ & $15452$
       & $\left(M_{\rm th} - 1.2\right)^{\dagger}_{-2.9}$
       & $\left(M_{\rm th} - 10\right)^{+9}_{-13}$ \\
       $\Xi_{bb}^*\bar{B}$ & $1({\frac{3}{2}}^-)$ & $C_{1a}$ & $15430$
       & $\left(M_{\rm th} - 0.3\right)^{\dagger}_{-2.4}$
       & $\left(M_{\rm th} - 12\right)^{+11}_{-13}$ \\
       $\Xi_{bb}^*\bar{B}^*$ & $1({\frac{1}{2}}^-)$
       & $C_{1a}-\frac{5}{3}\,C_{1b}$ & $15476$
       & $\left(M_{\rm th} - 8\right)^{+8}_{-7}$
       & $\left(M_{\rm th} - 5\right)^{\dagger}_{-8}$ \\
       $\Xi_{bb}^*\bar{B}^*$ & $1({\frac{3}{2}}^-)$ & 
$C_{1a}-\frac{2}{3}\,C_{1b}$ & $15476$
       & $\left(M_{\rm th} - 2.5\right)^{\dagger}_{-3.6}$
       & $\left(M_{\rm th} -9\right)^{+9}_{-11}$ \\
       $\Xi_{bb}^*\bar{B}^*$ & $1({\frac{5}{2}}^-)$ & $C_{1a}+C_{1b}$ & $15476$
       & $\left(M_{\rm th}-4.3\right)$[V]$_{+3.3}^{\rm N/A}$
       & $\left(M_{\rm th} - 18\right)^{+17}_{-19}$ \\
   \end{tabular}
\end{table}

The existence of  heavy meson-antimeson molecules implies the possibility of  partners
composed of a heavy meson and a doubly-heavy baryon (triply-heavy
pentaquarks)~\cite{Guo:2013xga}. This is based on the approximate heavy
antiquark-diquark symmetry (HADS) that emerges
from the observation that the interactions of heavy color triplet
objects with the light degrees of freedom (quarks and gluons) are
independent of the heavy color triplet's  spin and mass~\cite{Savage:1990di}.  

A first  consequence is that the spectrum of baryons with two heavy quarks can be related to the
spectrum of heavy light mesons with the same light degree of freedom
quantum numbers~\cite{Savage:1990di}.  The heavy diquark component of the baryon forms a color anti-triplet
with a characteristic length scale of $1 / (m_Q v)$,
where $m_Q$ is the mass of the heavy quarks
and $v$ their velocity. The length scale of the diquark is smaller than the typical QCD
length scale $1/\Lambda_{\rm QCD}$ and hence we can treat
the diquark as point-like if the quarks are heavy enough.

Within this scheme, we hint the existence of several baryonic partners of the
$X(3872)$ with isospin $I=0$ and $J^P = {\frac{5}{2}}^-$ or
$\frac32^-$. Moreover, we predict various $\Xi_{bb}^* \bar{B}^*$
triply-heavy pentaquarks with quantum numbers $I(J^P) =
1({\frac{1}{2}}^-)$ and $I(J^P) = 1({\frac{3}{2}}^-)$ partners of the
$Z_b(10610)$ $B^*\bar{B}^*$ molecule.  

We compile  some of these approximate predictions from \cite{Guo:2013sya} in
Table~\ref{tab:2}. They are subject to larger uncertainties than those
collected in Table~\ref{tab:1}, since violations of HADS are expected
to be larger than HQSS and HFS ones. i.e., ${\cal O} (\Lambda_{\rm QCD} / (m_Q v))$ vs ${\cal O} (\Lambda_{\rm QCD} / m_Q)$.

\section{The $\X \to \DD\pi^0$ decay and the long-distance structure
  of the $X(3872)$ resonance}

\begin{figure*}
  \includegraphics[width=1.0\textwidth]{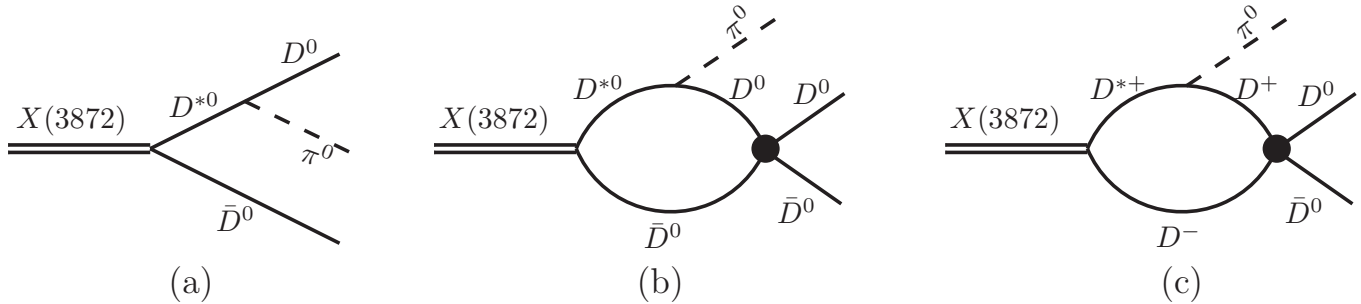}
\caption{Feynman diagrams for the decay
  $\X\to\DD\pi^0$. The charge conjugate
channel is not shown but included in the calculations.}
\label{fig:1}       
\end{figure*}

Within the molecular picture of  the $\X$ resonance, in its decay modes    with a
charmonium in the final state ($J/\psi\pi\pi$, $J/\psi3\pi$,
$J/\psi\gamma$ and $\psi'\gamma$), the heavy  quarks of the $D\bar D^*$
meson pair have to recombine to form the final charmonium. As a
consequence, these processes are not sensitive to
the $D\bar D^*$ wave function at long distances 
which is governed by the binding energy, but rather they are
determined by the short distance part of the $\X$ wave-function~\cite{Gamermann:2009uq}. 

The transition from the charm--anti-charm meson pair into the $J/\psi$
plus pions (or a photon), occurs at a distance much smaller than both
the size of the $\X$ as a hadronic molecule and  the range of
forces between the $D$ and $\bar D^*$ mesons. However, in the case of
the $\X \to \DD\pi^0$ decay, one  of the constituent hadrons ($D^0$)
is in the 
final state and the rest of the final particles are products of the decay of 
the other constituent hadron ($\bar D^{*0}$) of the $\X$
molecule. Thus, in this decay the relative
distance between the $D\bar D^*$ mesons can be as large as
allowed by the size of the $\X$ resonance, since the final state is
produced by the decay of the $\bar D^*$ meson instead of a
rescattering transition. Actually, it can be proved that within some 
approximations, the $d\Gamma/d|\vec{p}_{D^0}|$ distribution 
is related to the $\X$ wave-function $\Psi(\vec{p}_{D^0})$ \cite{Guo:2014hqa}.

\begin{figure*}
  \includegraphics[width=0.48\textwidth]{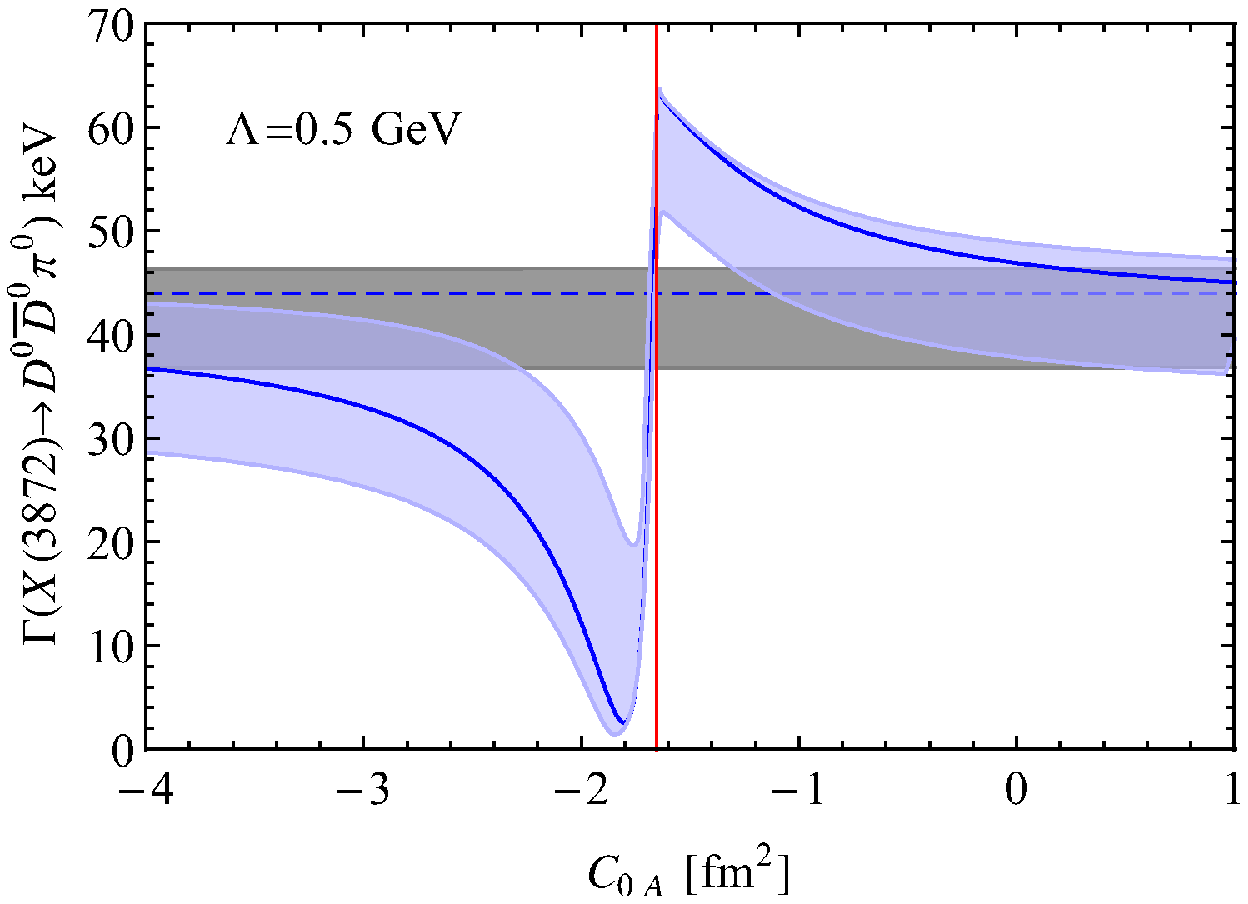}\hspace{0.5cm}\includegraphics[width=0.48\textwidth]{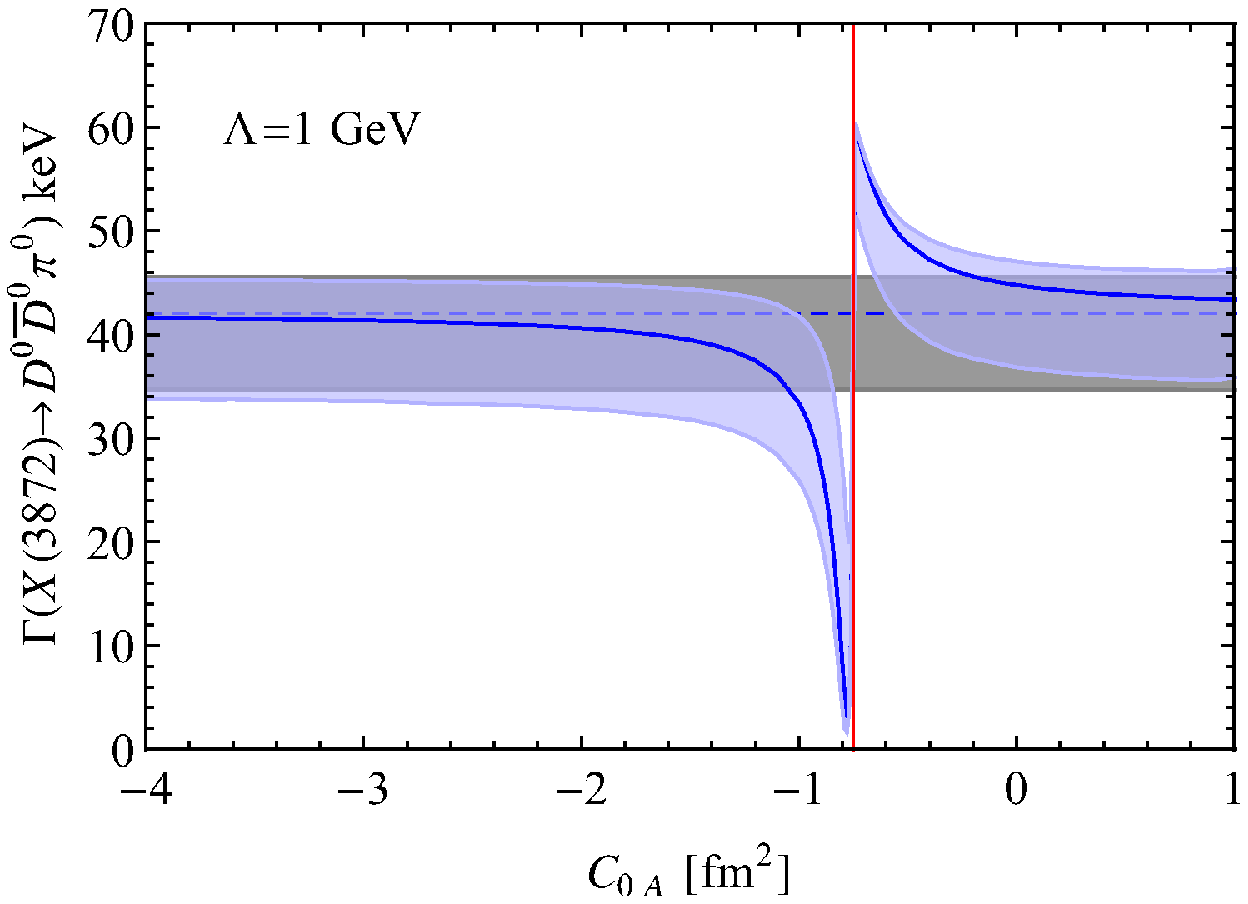}%
\caption{$\X\to D^0\bar D^0\pi^0$ partial decay
      width  as a function of  $C_{0A}$. The UV cutoff is set
      to $\Lambda=0.5$ GeV (1 GeV) in the left (right) panel.  The
      blue error bands contain $D\bar D$ FSI effects, while the grey
      bands stand for the tree level predictions (see Ref~\cite{Guo:2014hqa} for details).}
\label{fig:2}       
\end{figure*}

We have estimated the $\X \to \DD\pi^0$ decay width by evaluating the
diagrams depicted in Fig.~\ref{fig:1}. The tree level contribution
is fully determined by the $D^0\bar D^{*0}\pi$ coupling ($g/f_\pi$), the $\X$ mass
and its coupling constant to the neutral $D^0\bar D^{*0}$ channel ($g_0^X$),
which is determined by the residue of the $T-$matrix element at the
$\X$ pole. We find~\cite{Guo:2014hqa}
\begin{equation}
\label{eq:TreeLevel}
T_{\rm tree} = - 2i \frac{gg_0^X}{f_{\pi}}
\sqrt{M_{X}} M_{D^{*0}} M_{D^0} \vec{\epsilon}_{X} \cdot 
\vec{p}_{\pi} \left( \frac{1}{p_{12}^{2} - M_{D^{*0}}^{2}} + 
\frac{1}{p_{13}^{2} - M_{D^{*0}}^{2}} \right),
\end{equation}
where $\vec\epsilon_X$ is the polarization vector of the $\X$, $\vec
p_\pi$ is the three-momentum of the pion, $p_{12}$ and $p_{13}$ are
the four momenta of the $\pi^0D^0$ and $\pi^0\bar D^0$ systems,
respectively. Taking into account the phase space,
\begin{equation}
d\Gamma = \frac{1}{(2\pi)^3}\frac{1}{32 M_X^3} \overline{|T|}^2
dm_{12}^2 dm_{23}^2
\end{equation}
with the invariant masses $m_{12}^2 = p_{12}^2$ and $m_{23}^2=
(M^2_X+m^2_{\pi^0}+2M_{D^0}^2-m_{12}^2-p_{13}^2)$   of the final
$\pi^0D^0$ and $\DD$ pairs,  we readily obtain
\begin{eqnarray}
\Gamma_{\rm tree} &=& \frac{g^2}{192\pi^3 f_\pi^2}\left (g_0^X \frac{M_{D^0}
  M_{D^{*0}}}{M_X} \right)^2 \int^{(M_X-M_{D^0})^2}_{(M_{D^0}+m_\pi^0)^2} dm_{12}^2\nonumber \\
&\times &
\int^{(m^2_{23})_{\rm (max)}}_{(m^2_{23})_{\rm (min)}}d m^2_{23}\left( \frac{1}{p_{12}^{2} - M_{D^{*0}}^{2}} + 
\frac{1}{p_{13}^{2} - M_{D^{*0}}^{2}} \right)^2
|\vec{p}_\pi|^2 
\end{eqnarray}
where $ |\vec{p}_\pi| =\lambda^{1/2}(M_X^2,m^2_{23},m_{\pi^0}^2)/2M_X$ is the pion  momentum in the $\X$  center of mass frame [
$\lambda(x,y,z)=x^2+y^2+z^2-2(xy+yz+xz)$]. In addition, for a given value of $m^2_{12}$, the range of $m^2_{23}$ is determined by:
\begin{eqnarray}
(m^2_{23})_{\rm (max,min)} & = & (E^*_D+E^*_{\bar D})^2 -(p^*_D\mp p^*_{\bar
    D})^2  
\end{eqnarray}
with $E^*_D = (m_{12}^2-m^2_{\pi^0}+M_{D^0}^2) /2m_{12}$ and
$E^*_{\bar D} =
(M^2_X-m_{12}^2-M_{D^0}^2) /2m_{12}$ the energies
of the $D^0$ and $\bar D^0$ in the $m_{12}$ rest frame, respectively, and
$p^*_{D,\bar D}$ the moduli of their corresponding three momenta. In Ref.~\cite{Guo:2014hqa}, we found 
\begin{equation}
\label{eq:restree}
    \Gamma(\X\to D^0\bar D^0\pi^0)_{\rm tree} = 
44.0_{-7.2}^{+2.4} 
\left( 42.0_{-7.3}^{+3.6} \right) ~{\rm keV},
\end{equation}
where the values outside and inside the
parentheses are obtained with UV Gaussian cutoffs of $\Lambda=0.5$ and
1~GeV, respectively, and the errors (grey bands in Fig.~\ref{fig:2}) 
reflect the uncertainty in the inputs ($M_{\X}$ and the ratio of decay
amplitudes for the $\X\to J/\psi\rho$ and $\X\to J/\psi\omega$ decays). 

Since we published these
results, new high precision measurements of the masses of the $D^0$
and $D^{*0}$ mesons have 
become available~\cite{Tomaradze:2014eha,Seth:2014}, which have led to a more
precise determination of the $\X$ binding energy,
$B=13\pm 192$ keV~\cite{Seth:2014}.  To obtain the central values and the errors of Eq.~(\ref{eq:restree})
and the 68\% confident level (CL) bands displayed in Fig.~\ref{fig:2}, we used $B=160\pm 170$ keV, and a Monte Carlo simulation was
performed to propagate errors\footnote{The
 error on the threshold energy $(M_{D^0}+M_{D^{*0}})$, $\sim$120 keV, was not taken into account  in
Ref.~\cite{Guo:2014hqa}.}. In the simulation, 
we rejected $\X$ binding energies values smaller than 10 keV, 
and those values were set to this minimum value. This
effectively amounts to consider $B=160^{+170}_{-150}$ keV, since 
 the Gaussian distribution of binding energies was truncated. We slightly
decreased the lower error to guaranty a bound state with a  
CL larger than 68\%, since the scheme followed in \cite{Guo:2014hqa}
only allows the computation of the width when the $\X$ state is bound.
This binding energy range leads to $g_0^X =
0.35^{+0.08}_{-0.18}\,(0.34^{+0.07}_{-0.18})~{\rm GeV}^{-1/2}$ (the
two values correspond to the $\Lambda=0.5$ and 1~GeV choices of the UV regulator), which is  not compatible with zero\footnote{In \cite{Guo:2014hqa}, 
larger lower errors for the coupling  $g_0^X =
0.35^{+0.08}_{-0.29}\,(0.34^{+0.07}_{-0.29})~{\rm GeV}^{-1/2}$ were quoted. These correspond to a minimum of the binding energy  ($\sim 0.1$ keV) 
much closer  to zero, though in the calculation of the width, the 10 keV cut, mentioned above, was used.}. 
Thus, the lower grey bands in Fig.~\ref{fig:1} turn out not
to be compatible with a zero width either. However,
the new determination of $B=13\pm 192$ keV makes much more probable
the very low  binding energies close to zero, or  an unbound resonance. In this case, we would like to point out   that the decay width
should decrease, and eventually should vanish,  as  the binding energy
approaches  zero. This is because for very small binding energies, all the couplings of a
  bound state tend to zero when the mass of the bound state gets closer to
  the lowest threshold~\cite{Weinberg:1965zz}. For the  case
  of the $\X$, this was re-derived in \cite{Gamermann:2009uq} and 
  explicitly shown that both the neutral $\X D^0\bar D^{*0}$ ($g_0^X$) and
  charged $\X D^+\bar D^{*-}$ couplings scale as $B^\frac14$. Because of the  
  quite limited phase space available in this $p-$wave decay, the decay
  width, however, increases very rapidly as the binding energy departs from zero.  In any case,
  the lower errors displayed in  Eq.~(\ref{eq:restree}) and Fig.~\ref{fig:2}
  should be now considered with some caution. Moreover, the effect of the $D^{*0}$ width, neglected in the present calculation,  
  becomes sizable for binding energies below 10 keV.

The last two diagrams in Fig.~\ref{fig:1} account for the $ D \bar{D}\to
D\bar D$ final state interaction (FSI) effects, which are considered
by means of the appropriated linear combinations of the isoscalar and
isovector $T-$matrices. To obtain these scattering amplitudes, the LO
contact potential, involving the four LEC's $C_{0A}$, $C_{0B}$, $C_{1A}$ and 
$C_{1B}$, is used~\cite{HidalgoDuque:2012pq,Guo:2014hqa}. As
commented,  the $\X$ and $Z_b(10610)$ inputs  determine only  three of
the four counter-terms. The value of  $C_{0A}$ is not fixed, and thus 
the $D\bar D$ FSI effects on this decay are not fully determined.  As can be
seen  in Fig.~\ref{fig:2}, these
effects might be quite large, because for a certain range of $C_{0A}$
values, a near-threshold isoscalar $D\bar D$ bound state could be
dynamically generated~\cite{Nieves:2012tt,HidalgoDuque:2012pq}. If in future experiments 
the partial decay width is measured, a significant  deviation from the
predicted tree level value   
will indicate a FSI  effect, which might  be used to
extract the value of $C_{0A}$.  However as discussed above, this could be obscured if
turned out that the actual binding energy of the $\X$ state is
smaller than let us say 10 keV.

%

%



\begin{acknowledgements}
C.~H.-D. thanks the support of the 
JAE-CSIC Program. This work is supported in 
part by the DFG and the NSFC through funds provided to the Sino-German CRC 110 
``Symmetries and the Emergence of Structure in QCD'', by the NSFC (Grant No. 
11165005), by the Spanish Ministerio de Econom\'\i a y Competitividad and 
European FEDER funds under the contract FIS2011-28853-C02-02 and the Spanish 
Consolider-Ingenio 2010 Programme CPAN (CSD2007-00042), by Generalitat 
Valenciana under contract PROMETEOII/2014/0068 and by the EU HadronPhysics3 
project, grant agreement no. 283286.

\end{acknowledgements}



\end{document}